Title page for the full length original research article:

# US Code growth 1991-2025

Christopher Mantzaris[1*] and Prof. Dr Ajda Fošner[1] (https://orcid.org/0000-0002-3590-4590)

[1] Faculty of Management, University of Primorska, Koper, 6101 Slovenia, European Union.

[*] Corresponding author's email: 68223065{at}student.upr.si

**100-word write up describing the importance of the work:**

Science is the discipline of advancing humanity's objective knowledge, particularly in areas where this is hard to do. Society needs to know the development of the laws that govern it, so it can counteract unsustainable or undesirable developments. Quantifying all US federal law is not only important for people in the US: The oldest democracy is an ideal case study demonstrating the world how even liberal-democratic governments tend to grow the corpus of regulations. Further, total US federal laws are so hard to quantify, that is has not been done in 15 years. This study introduces the objective, auditable and reproducible methodology to achieve just that.

# US Code growth 1991-2025

**Abstract**: This is the first scientific article since 2010 counting the words which are effective and permanent federal law in the United States (US) Code. The latest version of the US Code –published in 2025– is the largest since 1991, encompassing over 24.4 million words. The low since 1991 was 1993 at roughly 15 million words. The word count grew in 30 out of 33 years. The average characters per word –potentially indicative of complexity– grew even in all 33 years. The research also touches upon why it is undesirable for laws to be longer and longer and mentions possible remedies.

# INTRODUCTION

The responsibility of 'science' –or to avoid too generic terms: 'the academic world' or 'academic research'– is to find the truth where a journalistic reporting of the facts is impractical, for example because the truth is not easily accessible, such as hidden under complexity.

Examples where journalism is impractical include how a certain lifestyle or drug is related to certain symptoms or life spans; or whether certain organisational structures create more profitable companies than others, more satisfied customers, employees etc.

The goal of this academic research project is to find truth in regards to the development of US laws: How positive or negative has the growth been between 1991 and 2025?

A continuous positive growth of the total aggregate of US laws would indicate that the US legal system was on an unsustainable path. Why? A visualisation with a parable, a thought experiment, can help understand why laws should not grow limitlessly.

Why laws can not be infinitely long – A thought experiment: Laws are essentially the rules of society. One imagines a board game, in which society plays the game of trying to build a better life – through economic activity, resulting in wealth creation. Therefore, if the rulebook of that board game would contain dozens of millions of words, the game would be unplayable, because players would find themselves in an increasingly difficult challenge of playing the game without breaking any rules. In detail, the four reasons for that are:

1. Players would be increasingly busy with studying and understanding the rules and would have less time to actually play the game – that is, create a better life for themselves and others through economic activity resulting in wealth creation.

2. More rules means less opportunity for wealth creation, as they restrict possibilities.

3. What is more, society would need an increasing amount of 'referees', a larger and larger policing force, to enforce the rules. Otherwise, if society –or the players of the board game– would gain the impression that rules are not enforced, fewer and fewer board game players would respect the rules. As a result of that, chaos and anarchy become more likely – as opposed to orderly wealth creation.

4. Further, the referees or police forces do not directly contribute to wealth creation, but merely are necessary to maintain order. Therefore, less directly constructive players would be available, since more and more people were busy with merely enforcing the rules.

5. Lastly, the more rules there are, the more players would break rules. So, not only do more rules result in less personal freedom of the players in addition to all of the above, but players also need to be punished for breaking the rules, such as by skipping rounds – or in the real world: sitting in correctional facilities. That also means fewer players can create wealth.

This all illustrates that the ideal rules for society are a minimalistic rulebook: As short as possible and only as long and restrictive as absolutely necessary. In fact, one could argue every good written product ought to be as short as possible, in order to minimise waste of the reader's time.

The thought experiment also shows the importance of the research question: Society requires factual, objectively true information –ideally in numerical or 'quantitative' form– regarding the size of its rulebook. So that it can counteract negative developments or reinforce positive ones.

Other publications: Despite the importance, other publications that find quantitive data in regards to the length of total US laws are scarce. In fact, only one other publication attempted to count the words of all US federal law, namely: Bommarito II and Katz (2010) in "A mathematical approach to the study of the United States Code". They found that the US Code –which contains all US federal law, consolidated into a single written product– contains for 2008-2010 the following word counts (Bommarito II & Katz, 2010):

**Table 1**: Dates and word counts found by Bommarito II and Katz (2010)

| Time           | Word count  |
|----------------|-------------|
| 2008, October  | 22,823,405  |
| 2009, November | 23,919,248  |
| 2010, March    | 24,224,985  |

This hardly allows for a proper historical comparison. Yet, the numbers from this preexisting publication help verify whether the approach chosen in this research –described in the following section: "Materials and methods"– finds sensical numbers.

**MATERIALS AND METHODS**

**Materials**

After every calendar year, the US House of Representatives publishes the US Code that is valid starting with the new year. For example, the US Code 2009 is the US Code version, which contains all changes of the year 2009, is published in 2010 and effective starting 2010.

For the US Code 1991-2009, the US House of Representatives published .ISO-versions. Each ISO, in turn, contains a .SRC-file, which is a plain text file of the full US Code.

The versions 1994-2024 (2024 has been published in and is effective starting 2025, hence "1991-2025" in the title) are provided in .HTM-files. HTM files are text, which contains tags that format the plain text.

The links to all these files used can be found in the "Data availability" section. One of the challenging aspects is that both the HTM files as well as the ISOs' text files contain mostly text that is not even law in the narrower sense, see 'Methods'.

**Methods**

To analyse the respective files, two Python codes were created: One for the ISO's text file, (Python-ISO) and another for the HTM-files (Python-HTM), the links to which are placed in the "Data availability" section.

Both categories of files contain mostly text, that would not qualify as relevant words of the legal code. For example, both texts include information regarding which legislative acts enacted each law section, from when to when a section is effective and so on. Therefore, both Python codes use the following metrics when counting words

— For each law section, word and character counting starts with the headline of the section and ends with the last word of the law text of that section.

— Words are separated by either whitespace or long dashes (em dashes). For example 'It is—however—interesting to see' or 'In General.—Nothing in this Act' should each count as six words.

— Charts and other figures are not counted.

**Problems resulting from the file type differences, the 1991-1994 period and the transition thereafter**

If one were to only analyse 1994-2024, one would not need the ISO files nor the corresponding Python-ISO for it. However, including the ISOs yields a longer comparison, namely roughly since the end of the Cold War.

The issue: Due to the difference of the files, the two Python codes yield similar, yet slightly different results for their respective file types around 1994 – Yet, for the ISO's latest years –around 2007-2009– they yield almost 100% identical results.

To yield an improved comparability of the 1991-1993 (ISO) results with the 1994-2024 period (HTM), the Python-ISO results are adjusted with a multiplier, which corresponds to the overcount of the 1994-1996 ISO results. Therefore, the 1991-1993 Python-ISO results are divided by:

the word count of the Phython-ISO for 1994-1996

divided by

the word count of the Phython-HTM for 1994-1996;

which yields results better comparable to the rest of the HTM data.

This effectively extends the 1994-2024-results of the Python-HTM, by the 1991-1993-results of the Python-ISO. The "Results and discussion" section's Table 3 illustrates the extension. Table 3 –and Figure 1 thereafter– depict for 1991-1993 the results after applying the multiplier of 1/1.45 to the raw 1991-1993 ISO results, so what Table 3 contains is ≈69% of the raw ISO results for 1991-1993.

The methodology is verified fourfold:

1. The HTM results show a steady development of the word count, providing a realistic picture, without sharp drops or rises between years.

2. Manual audit: Randomised samples of US Code sections for different years and the Python results for them have been manually compared with the HTM results. The word count of the Python codes yielded the same results as the manual tests.

3. Other publications confirm the total word count: Bommarito II and Katz' (2010) results are in the same ballpark as the HTM and ISO results found here. They differ less than 18.5% from the HTM-results in all 3 years, see Table 2 below.

4. The two codes for both source files cross-validate each other: The two Python codes, analysing two different file types, both found word counts that are for the latest ISO-years, 2007-2009, always within 3.1% of each other.

**Table 2**: 2007-2009 HTM, ISO and Bommarito II and Katz (2010) result comparison

| Year | Word count results | | ISO HTM | Bommarito II and Katz (2010) | % of HTM |
|------|------|------|------|------|------|
| | HTM | ISO | | | |
| 2007 | 19,519,644 | 19,553,788 | 100.175% | 22,823,405 | 116.925% |
| 2008 | 20,188,975 | 19,580,762 | 96.987% | 23,919,248 | 118.477% |
| 2009 | 20,477,455 | 20,599,050 | 100.594% | 24,224,985 | 118.301% |

*Bommarito II and Katz' (2010) values correspond to each preceding year, since HTM and ISO results are year-end values, published in each respective following year.*

Adjustments: While the results fluctuate stronger from each other in some other years, the HTM-results are more authoritative, as they show less severe fluctuations and hence are likely closer to reality. Nonetheless, it should be noted that even the HTM-results are not necessarily a perfect, true word count. For example, possible inconsistencies in the way the US Code is presented in the HTM files might result in the Python code to count some sections incorrectly – even though no such consistencies were found in manual, random tests. The latter three confirmation tests mentioned above offer additional scrutiny and due diligence, yet an imperfect count could theoretically still be possible.

The following adjustments are made, to yield a single comprehensive result per year; for as many years as possible:

1. Word count: The 1994-1996 ISO word count divided by the 1994-1996 HTM word count is 1.45, if correctly rounded. Therefore, the 1991-1993 ISO results are each divided by 1.45, to adjust those ISO values to the HTM values. For the 1994-2024 periods, the HTM word counts are used.

2. Characters per word: The chosen characters per word for 1991-1993 are the ISO results, since there are no HTM values. For the 1994-2024 periods, the HTM values are used.

# RESULTS AND DISCUSSION

**Table 3**: Word count, characters per word and growth of both by year

| Year | Word count | Characters per word | *Growth of word count, in percent* | *Growth of characters per word, in permille* |
|---|---|---|---|---|
| 1991 | 18,447,706 | 6.09994 | - | - |
| 1992 | 17,048,645 | 6.11779 | *-7.58* | *2.93* |
| 1993 | 15,053,688 | 6.14434 | *-11.70* | *4.34* |
| 1994 | 15,357,213 | 6.25813 | *2.02* | *18.52* |
| 1995 | 15,747,251 | 6.26076 | *2.54* | *0.42* |
| 1996 | 16,162,383 | 6.26367 | *2.64* | *0.46* |
| 1997 | 16,424,834 | 6.26386 | *1.62* | *0.03* |
| 1998 | 16,676,295 | 6.26464 | *1.53* | *0.13* |
| 1999 | 16,836,511 | 6.26696 | *0.96* | *0.37* |
| 2000 | 17,308,741 | 6.27057 | *2.80* | *0.58* |
| 2001 | 17,516,122 | 6.27199 | *1.20* | *0.23* |
| 2002 | 17,938,097 | 6.27577 | *2.41* | *0.60* |
| 2003 | 18,277,526 | 6.27929 | *1.89* | *0.56* |
| 2004 | 18,579,468 | 6.28236 | *1.65* | *0.49* |
| 2005 | 18,957,299 | 6.28611 | *2.03* | *0.60* |
| 2006 | 19,179,051 | 6.29206 | *1.17* | *0.95* |
| 2007 | 19,519,644 | 6.29769 | *1.78* | *0.89* |
| 2008 | 20,188,975 | 6.30363 | *3.43* | *0.94* |
| 2009 | 20,477,455 | 6.30586 | *1.43* | *0.35* |
| 2010 | 21,190,616 | 6.31277 | *3.48* | *1.10* |
| 2011 | 21,231,510 | 6.31286 | *0.19* | *0.01* |
| 2012 | 21,445,800 | 6.31487 | *1.01* | *0.32* |
| 2013 | 21,510,087 | 6.31552 | *0.30* | *0.10* |
| 2014 | 21,811,339 | 6.31909 | *1.40* | *0.56* |
| 2015 | 21,671,536 | 6.32280 | *-0.64* | *0.59* |
| 2016 | 21,974,198 | 6.32916 | *1.40* | *1.01* |
| 2017 | 22,072,574 | 6.33054 | *0.45* | *0.22* |
| 2018 | 22,498,981 | 6.33507 | *1.93* | *0.72* |
| 2019 | 22,714,272 | 6.33695 | *0.96* | *0.30* |
| 2020 | 23,393,543 | 6.34386 | *2.99* | *1.09* |
| 2021 | 23,603,906 | 6.34600 | *0.90* | *0.34* |
| 2022 | 24,226,168 | 6.35316 | *2.64* | *1.13* |
| 2023 | 24,338,455 | 6.35388 | *0.46* | *0.11* |
| 2024 | 24,409,969 | 6.35444 | *0.29* | *0.09* |

1991-1993 values are based on the ISO files and associated Python code; 1994-2024 values are based on the HTM files and corresponding Python code. The 2024 file is the latest available version at the creation of this research, downloaded on 2025-01-04, as the 2024-archive version was not yet available at the time of the conducting of this research. It is not expected that the used and the archived version with a later publication date would have significantly different word counts. The links to all source files are provided in the Data availability section.

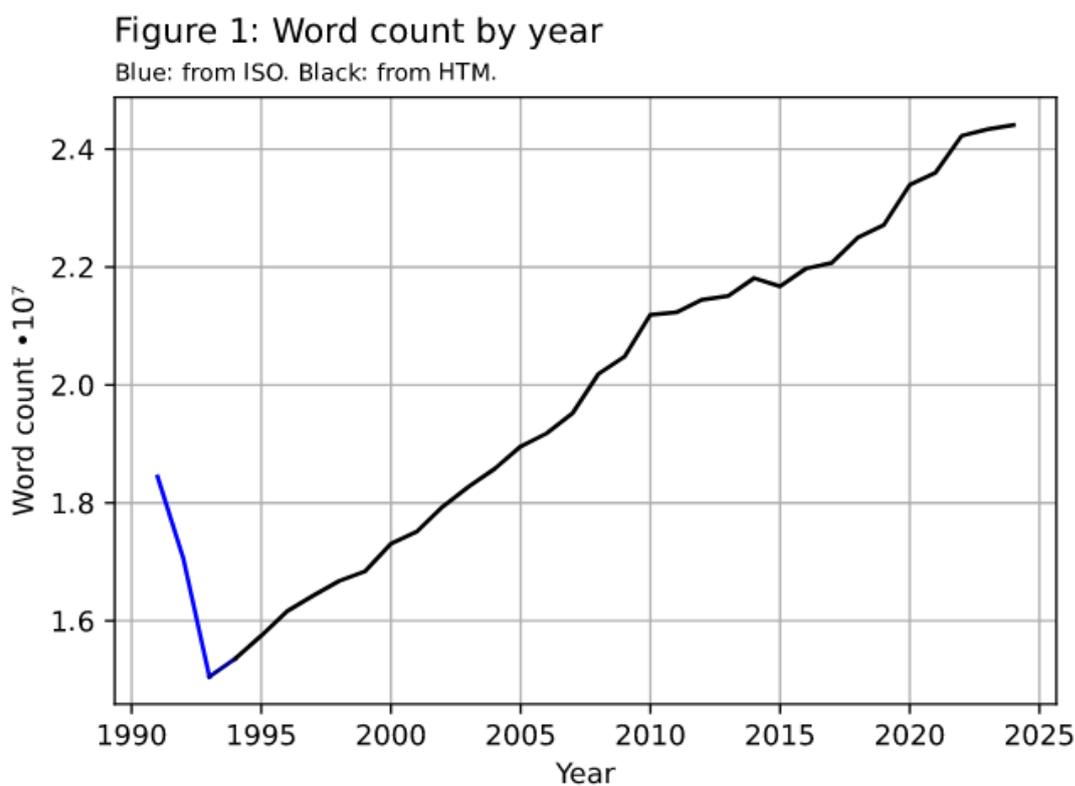

**Figure 1**: Word count by year

The ISO values (blue) have higher uncertainty, as they fluctuate for some years strongly – much more so than the HTM results. Yet regardless of how much weight one gives the blue ISO values, it does not distract from the clear upward trend since the early 1990s. The US Code's word count grew from roughly 15 million words in 1993 to over 24.4 million words in 2024.

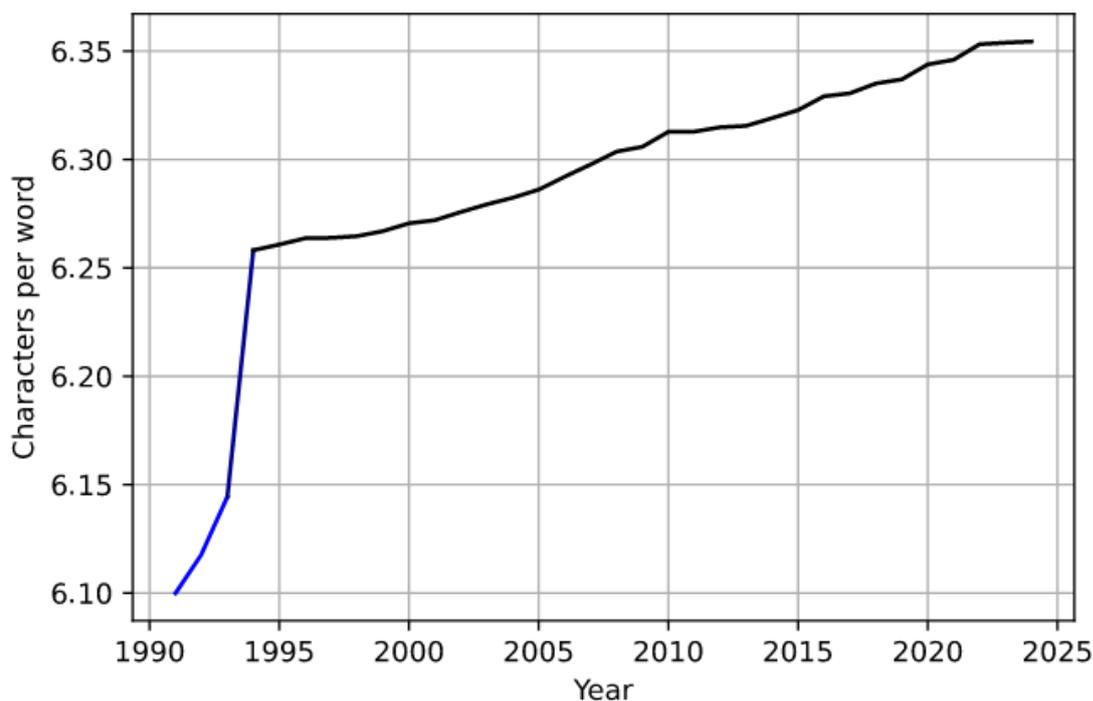

**Figure 2**: Characters per word by year

Some researchers –such as Katz and Bommarito II (2014, see under the chapter "Average Word Length" there)– believe that average word length is –generally speaking– an indicator for the complexity of (legal) text, and legal complexity is resource intensive and hence should be abandoned as much as possible – in favour of simplicity and thereby higher resource allocation for productive means. If one were to follow that premise and everything else being equal, then the results indicate that US Code complexity grew in every single year since 1991, because average characters per word did.

It should be noted that average characters per word is not equivalent to average letters per word, as the former includes whitespace –such as spaces or line breaks– and other non-letter characters, especially dashes. Logically, the average amount of whitespace characters per word is 1. The expected average letters per word are estimated to be the average characters per word, minus a value between 1.01 and 1.5. This is, because some words may be concatenated into a single word or split into two parts of a word with a dash, such as 'on-line' from 'on line' or 'online'. Also, other punctuation may be attached to a word, such as parentheses, commas, semicolons or full stops. The 1.01-1.5 estimation derives from the facts, that

– every character count of a word includes a (leading) whitespace (1), and

– at least a few words are connected with punctuation such as dashes for full stops (at least 1 + 0.01);

– but certainly not more than every second word (maximum 1 + 0.5).

**CONCLUSIONS**

US Code word count grew by 62.2% between 1993 and 2024, from roughly 15 million words to over 24.4 million words. In addition to the growth of the quantity of words, average characters per word grew in every single year in the observed 1991-2024 period. Some researchers understandably believe average characters per word growth indicates increased complexity, as simpler words are typically shorter – and vice versa.

An introductory explanation illustrated why limitless US Code word growth is unsustainable: Regulation burdens and limits society and infinite regulatory weight would be impossible or at least undesirable. Together with the results, policy and decision makers gain crucial information relevant for addressing the issue. A potential solution off the unsustainable path may be implementing possibilities for the electorate to directly revoke laws, such as via popular petitions. Yet another idea might include 'garbage collection' as known from computer science: Laws that have not been applied –or not have been applied enough– in, for example, 15 years should automatically become ineffective. It would merely require courts –and maybe also the executive branch or legal scholars– to report the laws they applied.

Other researchers are welcome to scrutinise or enhance the methodology –such as by improving the provided Python code or writing their own software– to reproduce, confirm or disconfirm the findings.

## ACKNOWLEDGEMENTS

The authors thank the peer-reviewers and the publication venue.

## DECLARATION OF CONFLICT OF INTEREST

The authors could not identify any potential conflicts of interest.

## FUNDING INFORMATION

No funding has been received for this research project.

## AUTHOR CONTRIBUTION

Conceptualization: C.M.; data collection: C.M.; statistical analysis: A.F., C.M.; data interpretation: A.F., C.M.; writing the first draft: C.M.; manuscript editing and reviewing: A.F., C.M.; project administration: A.F.

All authors have accepted responsibility for the content of the manuscript, reviewed all results, and approved the final version.

## DATA AVAILABILITY

**Input data 1 of 2 – ISO, archived via web.archive.org:**

web.archive.org/web/20130417072749if_/uscode.house.gov/iso/USC2009.iso

web.archive.org/web/20130417060641if_/uscode.house.gov/iso/USC2008.iso

web.archive.org/web/20130417051201if_/uscode.house.gov/iso/USC2007.iso

web.archive.org/web/20130417065225if_/uscode.house.gov/iso/USC2006.iso

web.archive.org/web/20130417034605if_/uscode.house.gov/iso/USC2005.iso

web.archive.org/web/20130417043444if_/uscode.house.gov/iso/USC2004.iso

web.archive.org/web/20120320042329if_/uscode.house.gov/iso/USC2003.iso

web.archive.org/web/20120320012555if_/uscode.house.gov/iso/USC2002.iso

web.archive.org/web/20120320093250if_/uscode.house.gov/iso/USC2001.iso

web.archive.org/web/20120320081916if_/uscode.house.gov/iso/USC2000.iso

web.archive.org/web/20120320024047if_/uscode.house.gov/iso/USC1999.iso

web.archive.org/web/20120320053844if_/uscode.house.gov/iso/USC1998.iso

web.archive.org/web/20120320040500if_/uscode.house.gov/iso/USC1997.iso

web.archive.org/web/20120320020930if_/uscode.house.gov/iso/USC1996.iso

web.archive.org/web/20120320102806if_/uscode.house.gov/iso/USC1995.iso

web.archive.org/web/20070725221137if_/uscode.house.gov/iso/USC1994.iso

web.archive.org/web/20070725211344if_/uscode.house.gov/iso/USC1993.iso

web.archive.org/web/20090805131959if_/uscode.house.gov/iso/USC1992.iso

web.archive.org/web/20060830225943if_/uscode.house.gov/iso/USC1991.iso

Unfortunately unable to download or find:

web.archive.org/web/20060726195741/uscode.house.gov/iso/USC1990.iso

or anything earlier.

**Input data 2 of 2 – HTM, archived via web.archive.org:**

web.archive.org/web/20241125194950if_/uscode.house.gov/download/annualhistoricalarchives/XHTML/1994.zip

web.archive.org/web/20241125194936if_/uscode.house.gov/download/annualhistoricalarchives/XHTML/1995.zip

web.archive.org/web/20241125194923if_/uscode.house.gov/download/annualhistoricalarchives/XHTML/1996.zip

web.archive.org/web/20241125194909if_/uscode.house.gov/download/annualhistoricalarchives/XHTML/1997.zip

web.archive.org/web/20241125194845if_/uscode.house.gov/download/annualhistoricalarchives/XHTML/1998.zip

web.archive.org/web/20241125194829if_/uscode.house.gov/download/annualhistoricalarchives/XHTML/1999.zip

web.archive.org/web/20241125194809if_/uscode.house.gov/download/annualhistoricalarchives/XHTML/2000.zip

web.archive.org/web/20241125194754if_/uscode.house.gov/download/annualhistoricalarchives/XHTML/2001.zip

web.archive.org/web/20241125194732if_/uscode.house.gov/download/annualhistoricalarchives/XHTML/2002.zip

web.archive.org/web/20241125194717if_/uscode.house.gov/download/annualhistoricalarchives/XHTML/2003.zip

web.archive.org/web/20241125194654if_/uscode.house.gov/download/annualhistoricalarchives/XHTML/2004.zip

web.archive.org/web/20241125194630if_/uscode.house.gov/download/annualhistoricalarchives/XHTML/2005.zip

web.archive.org/web/20241125194610if_/uscode.house.gov/download/annualhistoricalarchives/XHTML/2006.zip

web.archive.org/web/20241125194556if_/uscode.house.gov/download/annualhistoricalarchives/XHTML/2007.zip

web.archive.org/web/20241125194542if_/uscode.house.gov/download/annualhistoricalarchives/XHTML/2008.zip

web.archive.org/web/20241125194526if_/uscode.house.gov/download/annualhistoricalarchives/XHTML/2009.zip

web.archive.org/web/20241125194502if_/uscode.house.gov/download/annualhistoricalarchives/XHTML/2010.zip

web.archive.org/web/20241125194447if_/uscode.house.gov/download/annualhistoricalarchives/XHTML/2011.zip

web.archive.org/web/20241125194423if_/uscode.house.gov/download/annualhistoricalarchives/XHTML/2012.zip

web.archive.org/web/20241125194359if_/uscode.house.gov/download/annualhistoricalarchives/XHTML/2013.zip

web.archive.org/web/20241125194338if_/uscode.house.gov/download/annualhistoricalarchives/XHTML/2014.zip

web.archive.org/web/20241125194310if_/uscode.house.gov/download/annualhistoricalarchives/XHTML/2015.zip

web.archive.org/web/20241125194255if_/uscode.house.gov/download/annualhistoricalarchives/XHTML/2016.zip

web.archive.org/web/20241125194233if_/uscode.house.gov/download/annualhistoricalarchives/XHTML/2017.zip

web.archive.org/web/20241125194136if_/uscode.house.gov/download/annualhistoricalarchives/XHTML/2018.zip

web.archive.org/web/20241125194122if_/uscode.house.gov/download/annualhistoricalarchives/XHTML/2019.zip

web.archive.org/web/20241125194049if_/uscode.house.gov/download/annualhistoricalarchives/XHTML/2020.zip

web.archive.org/web/20241125194026if_/uscode.house.gov/download/annualhistoricalarchives/XHTML/2021.zip

web.archive.org/web/20241125193951if_/uscode.house.gov/download/annualhistoricalarchives/XHTML/2022.zip

web.archive.org/web/20241125193930if_/uscode.house.gov/download/annualhistoricalarchives/XHTML/2023.zip

web.archive.org/web/20250104/uscode.house.gov/download/releasepoints/us/pl/118/158/htm_uscAll@118-158.zip

**Python code to analyse input data 1 of 2 – For ISO:**

pastejustit.com/raw/2025-01pythoncodeisoforacademicpublicationbycm-af

– archived on 2025-01-09 at archive.md/jSeoX.

**Python code to analyse input data 2 of 2 – For HTM:**

pastejustit.com/raw/2025-01pythoncodehtmforacademicpublicationbycm-af

– archived on 2025-01-09 at archive.md/EqJyt.

**Consolidated results data, as a tab separated value file:**

pastejustit.com/raw/2025-01pythonconsolidatedresultsastsvforacademicpublicationbycm-af

– archived on 2025-01-09 at archive.md/LHBae.

**REFERENCES**

Bommarito II, M. J., & Katz, D. M. (2010). A mathematical approach to the study of the United States Code. *Physica A: Statistical Mechanics and its Applications*, *389*(19), 4195-4200. arxiv.org/pdf/1003.4146 or doi.org/10.1016/j.physa.2010.05.057

Katz, D. M., & Bommarito II, M. J. (2014). Measuring the complexity of the law: the United States Code. *Artificial intelligence and law, 22*, 337-374. doi.org/10.2139/ssrn.2307352 or doi.org/10.1007/s10506-014-9160-8

**TABLES**

**Table 1**: Dates and word counts found by Bommarito II and Katz (2010)

| Time | Word count |
|---|---|
| 2008, October | 22,823,405 |
| 2009, November | 23,919,248 |
| 2010, March | 24,224,985 |

**Table 2**: 2007-2009 HTM, ISO and Bommarito II and Katz (2010) result comparison

| Year | Word count results | | ISO/HTM | *Bommarito II and Katz (2010)* | *% of HTM* |
|---|---|---|---|---|---|
| | **HTM** | **ISO** | | | |
| 2007 | 19,519,644 | 19,553,788 | 100.175% | *22,823,405* | *116.925%* |
| 2008 | 20,188,975 | 19,580,762 | 96.987% | *23,919,248* | *118.477%* |
| 2009 | 20,477,455 | 20,599,050 | 100.594% | *24,224,985* | *118.301%* |

Table 3: Word count, characters per word and growth of both by year

| Year | Word count | Characters per word | Growth of word count, in percent | Growth of characters per word, in permille |
|---|---|---|---|---|
| 1991 | 18,447,706 | 6.09994 | - | - |
| 1992 | 17,048,645 | 6.11779 | -7.58 | 2.93 |
| 1993 | 15,053,688 | 6.14434 | -11.70 | 4.34 |
| 1994 | 15,357,213 | 6.25813 | 2.02 | 18.52 |
| 1995 | 15,747,251 | 6.26076 | 2.54 | 0.42 |
| 1996 | 16,162,383 | 6.26367 | 2.64 | 0.46 |
| 1997 | 16,424,834 | 6.26386 | 1.62 | 0.03 |
| 1998 | 16,676,295 | 6.26464 | 1.53 | 0.13 |
| 1999 | 16,836,511 | 6.26696 | 0.96 | 0.37 |
| 2000 | 17,308,741 | 6.27057 | 2.80 | 0.58 |
| 2001 | 17,516,122 | 6.27199 | 1.20 | 0.23 |
| 2002 | 17,938,097 | 6.27577 | 2.41 | 0.60 |
| 2003 | 18,277,526 | 6.27929 | 1.89 | 0.56 |
| 2004 | 18,579,468 | 6.28236 | 1.65 | 0.49 |
| 2005 | 18,957,299 | 6.28611 | 2.03 | 0.60 |
| 2006 | 19,179,051 | 6.29206 | 1.17 | 0.95 |
| 2007 | 19,519,644 | 6.29769 | 1.78 | 0.89 |
| 2008 | 20,188,975 | 6.30363 | 3.43 | 0.94 |
| 2009 | 20,477,455 | 6.30586 | 1.43 | 0.35 |
| 2010 | 21,190,616 | 6.31277 | 3.48 | 1.10 |
| 2011 | 21,231,510 | 6.31286 | 0.19 | 0.01 |
| 2012 | 21,445,800 | 6.31487 | 1.01 | 0.32 |
| 2013 | 21,510,087 | 6.31552 | 0.30 | 0.10 |
| 2014 | 21,811,339 | 6.31909 | 1.40 | 0.56 |
| 2015 | 21,671,536 | 6.32280 | -0.64 | 0.59 |
| 2016 | 21,974,198 | 6.32916 | 1.40 | 1.01 |
| 2017 | 22,072,574 | 6.33054 | 0.45 | 0.22 |
| 2018 | 22,498,981 | 6.33507 | 1.93 | 0.72 |
| 2019 | 22,714,272 | 6.33695 | 0.96 | 0.30 |
| 2020 | 23,393,543 | 6.34386 | 2.99 | 1.09 |
| 2021 | 23,603,906 | 6.34600 | 0.90 | 0.34 |
| 2022 | 24,226,168 | 6.35316 | 2.64 | 1.13 |
| 2023 | 24,338,455 | 6.35388 | 0.46 | 0.11 |
| 2024 | 24,409,969 | 6.35444 | 0.29 | 0.09 |

**FIGURES**

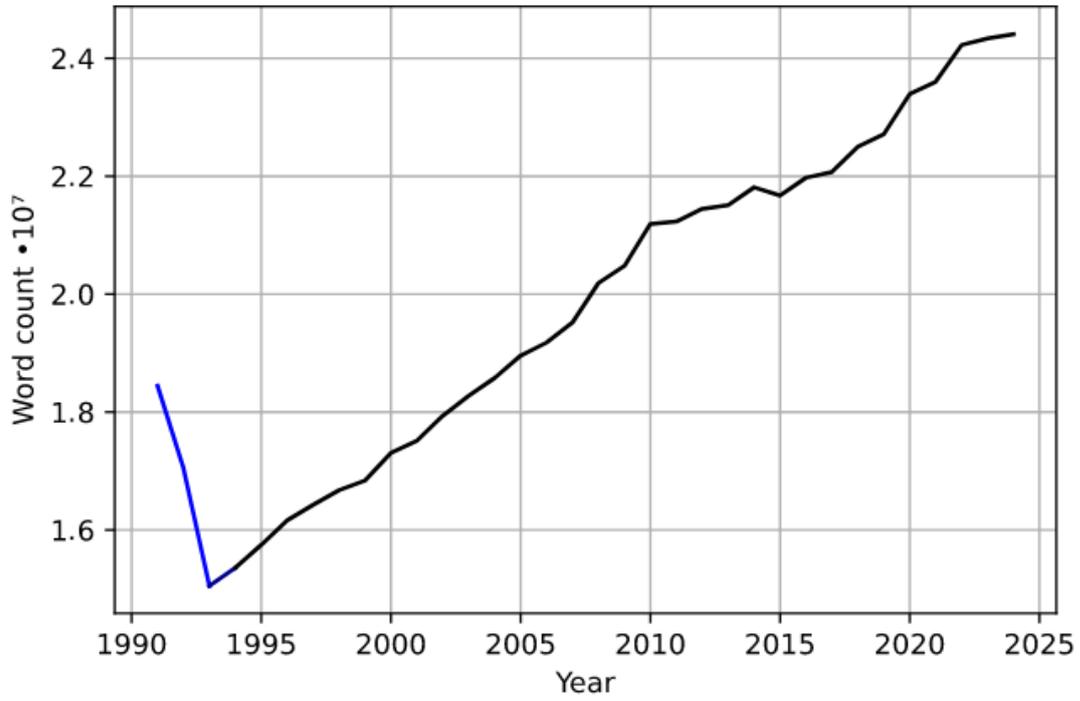

**Figure 1**: Word count by year

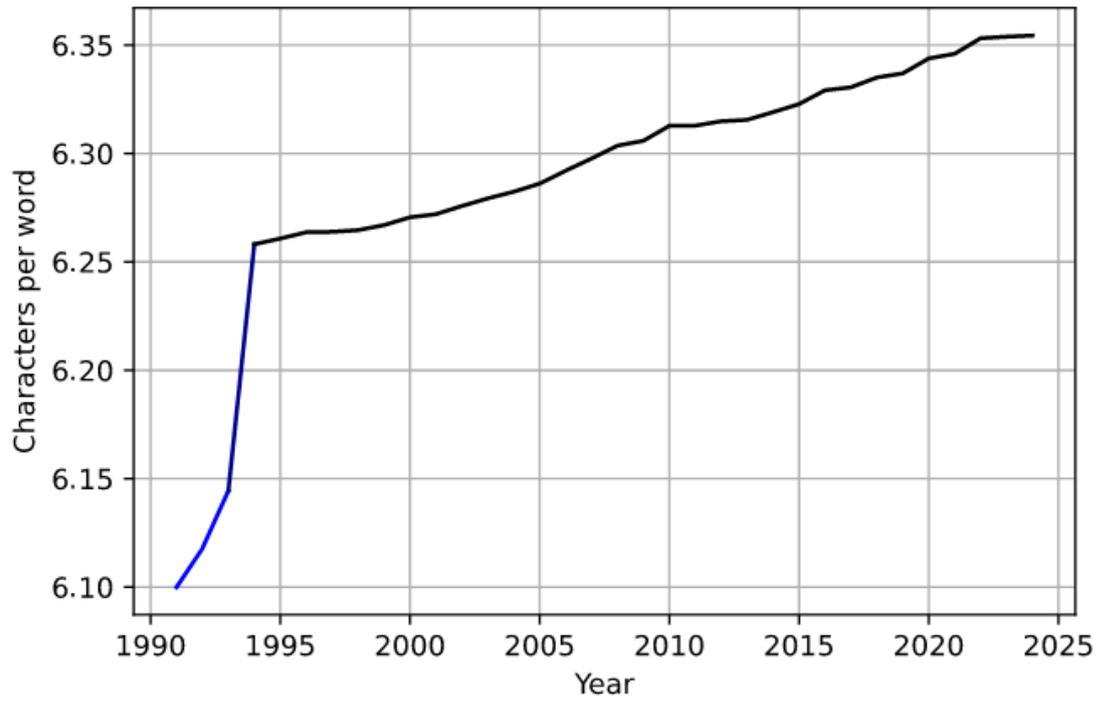

**Figure 2**: Characters per word by year